\documentclass[10pt,paper]{IEEEtran}
\usepackage{mathrsfs}
\usepackage{amsfonts}
\usepackage[dvips]{graphics,color}
\usepackage{float, tikz, array, caption, subcaption}
\usepackage{algorithm,algorithmic}
\usepackage{latexsym,amssymb}
\usepackage{amsmath, amsbsy}
\usepackage{amsopn,amstext}
\usepackage{cite}

\newtheorem{theorem}{Theorem}
\newtheorem{definition}{Definition}
\newtheorem{lemma}{Lemma}

\def\proof{\noindent{\bf Proof: }}

\newcommand{\be}{\begin{equation}}
\newcommand{\ee}{\end{equation}}

\newcommand{\bea}{\begin{eqnarray}}
\newcommand{\eea}{\end{eqnarray}}

\def\ba{\begin{array}}
\def\ea{\end{array}}

\begin{document}
\title{Opportunistic Wireless Control Over State-Dependent Fading Channels}

\author{{Shuling Wang, Peizhe Li, Shanying Zhu and Cailian Chen}
\thanks{This work was supported by National Key
R\&D Program of China under the grant 2018YFB1703201,
the NSF of China under the grants 61922058, 62173225,
and Chinese Ministry of Education Research Found on Intelligent Manufacturing
under the grant MCM20180703.}
 \thanks{The authors are with the Department of Automation, Shanghai Jiao Tong University,
Shanghai 200240, China;
Key Laboratory of System Control and Information Processing, Ministry of Education of China,
Shanghai 200240, China, and also Shanghai Engineering Research Center of Intelligent Control
and Management, Shanghai 200240, China. E-mails: shulingwang2021\_sjtu@sjtu.edu.cn, lipeizhe2020@sjtu.edu.cn, shyzhu@sjtu.edu.cn, cailianchen@sjtu.edu.cn.}}

%\markboth{IEEE TRANSACTIONS ON NEURAL NETWORKS AND LEARNING SYSTEMS}%
%{Wang \MakeLowercase{\textit{et al.}}: Graph-Based Function Perturbation Analysis for Observability of Multi-Valued Logical Networks} %\vspace*{1\baselineskip}

 \maketitle

 \begin{abstract}
The heterogeneous system consisting of the wireless control system (WCS) and
mobile agent system (MAS) is ubiquitous in Industrial Internet of Things (IIoT) systems.
Within this system, the positions of mobile agents may lead to shadow
fading on the wireless channel that the WCS is controlled over and can significantly
compromise the WCS's performance. This paper focuses on the controller design for the MAS
to ensure the performance of WCS in the presence of WCS and MAS coupling. Firstly,
the constrained finite field network (FFN) with profile-dependent switching topology is
adopted to proceed the operational control for the MAS. By virtue of the algebraic state
space representation (ASSR) method, an equivalent form is obtained for the WCS and MAS
coupling. A necessary and sufficient condition in terms of constrained set stabilization
is then established to ensure the Lyapunov-like performance with expected decay
rate.
Finally, a graphical method together with the breath-first searching is provided to design
state feedback controllers for the MAS. With this method, it is easy to check the
constrained set stabilization of MAS and to ensure
the performance requirements of WCS in the presence of WCS and MAS coupling.
The study of an illustrative example shows the effectiveness of the proposed method.
 \end{abstract}

% \begin{IEEEkeywords}
%Multi-valued logical network, Observability, Function perturbation, Transition graph, Algebraic state space representation.
%  \end{IEEEkeywords}

\section{Introduction}
Internet of Things (IoT) is an emerging domain in which all things are interconnected to
realize dynamic information interaction \cite{fuqaha2015}. Restricting the things in IoT
to the industrial scenario, Industrial Internet of Things (IIoT) enables flexible, efficient
and sustainable production in many fields including smart manufacturing through numerous plants that can be reconfigured based on
process requirements \cite{Baumann2021,fadlullah2011}. These plants consist
of heterogeneous physical systems with sensing, computation, communication and actuation
capabilities, which interact autonomously with each other by exchanging
information over wireless networks \cite{vitturi2019}.

In typical IIoT applications, e.g., smart manufacturing, heterogeneous physical systems which may have different objectives
coordinate with each other to jointly perform overall task \cite{Baumann2021,Ahlen2019}.
The typical characteristic of the IIoT system lies in that numerous wireless sensors are
deployed to monitor and control industrial plants forming wireless control networks (WCSs).
An integral aspect of IIoT in smart manufacturing are mobile agents, which coordinate with
the WCS in various ways, and can flexibly execute diverse tasks
%\textbf{such as autonomously visually inspect and monitor plants, sense information and transport
%objects or personnel}
contributing to the overall
manufacturing process.
%Recently, mobile agents have been playing an increasingly important role in IIoT systems to
%\cite{He2022,Pulikottil2021}.
Mobile agents system (MAS)
can leverage the full potential of mobile agents to autonomously visually inspect and
monitor plants, transport objects or personnel, and jointly carry complex and large
components \cite{He2022,Pulikottil2021,liang2019}.
Such coordination of heterogeneous systems is necessary, e.g., manufacturing
systems with heavy machines and cranes \cite{Agrawal2014}, assembly
processes with autonomous assembly arms and forklifts \cite{hu2019}
and sensor networks with mobile agents \cite{Quevedo2013}.
%Several studies have been conducted on collaborative control of MASs
%such as
%formation control \cite{tanner2003}, path planning \cite{Choset2005}.
However, the joint coordination of MASs and WCSs brings heterogeneity to IIoT systems.
Moreover, the positions of mobile agents may lead to shadow fading on the wireless channel
that the
WCS is controlled over and can significantly compromise the WCS's
performance. The
fading channel measurements acquired at a rolling mill at
Sandvik in Sweden \cite{Agrawal2014} show that mobile machinery
and cranes in the ceiling leads to substantial variations in
the measured channel gains. An industrial situation where
different positions of mobile robots lead to different fading
distributions is characterized in \cite{Quevedo2013}.
Therefore, it is necessary to explicitly examine the state-dependency of wireless channels
due to such WCS
and MAS coupling to ensure the performance of WCSs.

The channel gains used to characterize shadow fading are traditionally modeled as
independent identical distributed random processes \cite{Gatsis2014} or Markov chains
\cite{zhangq1999}, where the network state is assumed to be independent from
physical states. Under this assumption, research on scheduling wireless
network parameters has been conducted to satisfy given control performance requirements
such as stability \cite{gatsis2015,gatsis2018,Hristu2001,Molin}, controllability and
observability \cite{zhang2006} and minimizing linear quadratic objectives \cite{le2011}.
A survey on design and optimization for WCSs is presented in
\cite{Park2018}. It is obvious that these methods on performance analysis of WCS
fail in the scenario where the WCS and MAS coupling exists.
Few works addressing the dependence of network state on physical states have been
conducted.
Reference \cite{hu2020} considers the power control in vehicular WCS, where the influence of
vehicles's physical states on the network state has been established to
ensure the performance of the WCS.
In \cite{hu2019}, the authors study a state-dependent channel model, based on which a novel
co-design paradigm addressing the
coupling between a WCS and a single mobile agent has been
 proposed to achieve
safety and efficiency of overall IIoT system. A Markov decision process is adopted to
characterize the dynamics of a mobile agent, which is, however, limited to the single
agent case.

In this paper, we consider a heterogeneous IIoT system where a WCS and an MAS
coordinate with each other to jointly perform overall task over state-dependent fading channels.
We focus on the opportunistic wireless control to ensure the Lyapunov-like performance with expected decay
rate of the WCS in the presence of WCS and MAS coupling.
How to convert
the performance requirements of the WCS to a specific control objective of MAS is the
key to this problem.

The contributions of this paper are two-fold. Firstly, by discretizing the workshop of the
MAS into finite two dimensional regions, the constrained finite field network (FFN) with
profile-dependent switching
topology is adopted to proceed the
operational control for the MAS, which requires finite communication, memory and computation
resources, and can achieve finite-time convergence. To the best of our knowledge, there are
few results on the analysis and control of constrained FFNs with profile-dependent switching
topology. Secondly, based on the algebraic state
space representation (ASSR) method,
the performance requirements of the WCS is converted into the constrained set stabilization of
the FFN.
A graphical method together with the breath-first searching is proposed for the controller
design for the MAS to ensure the performance requirements of WCS in the presence of
WCS and MAS coupling.
Compared with the existing results
on set stabilization of finite-value systems obtained via algebraic
approaches \cite{guo2015,lif2017,zhangq2021}, the results
proposed in this paper is more computationally economical.

This paper is organized as follows. Section II characterizes the heterogeneous IIoT system
and problem formulation. Lyapunov-like performance analysis of the WCS is presented in Section III. In Section IV, main results are
demonstrated by an illustrative example, which is followed by the conclusion in Section V.

Notations: The cone of $m\times m$ real positive
definite (semi-definite) matrices is denoted by $S_{++}^m$ ($S_{+}^m$). Denote the $s$-th
column and the $s$-th row of matrix $F$ by $Col_s(F)$ and $Row_s(F)$, respectively.
For $s\in\mathbb{Z}_+$, define $\mathcal{D}_s:=\{0,1,\cdots,s-1\}$ and $\Delta_s:=\{\delta_s^i: i=1, \cdots, s\}$,
where $\delta_s^i:=Col_i(I_s)$ and $I_s$ is the $s$-dimensional identity
matrix. ${\mathcal{L}}_{s\times t}$ consists of all $s\times t$ logical matrices in the form
of $M=[\delta_s^{i_1}\;\delta_s^{i_2}\;\cdots\;\delta_s^{i_t}]$, which is briefly expressed
as $M=\delta_s[i_1\;i_2\;\cdots\;i_t]$. Throughout this paper, semi-tensor product
($\ltimes$) is the basic matrix product \cite{cheng2011} defined as
$M\ltimes P:=(M\otimes I_{\frac{l}{n}})(P\otimes I_{\frac{l}{p}})$,
where $M\in{\mathbb{R}}_{m\times n}$, $P\in {\mathbb{R}}_{p\times q}$, $l$ is the least
common multiple of $n$ and $p$, and $\otimes$ is the Kronecker product. In most places of
this paper, the symbol ``$\ltimes$'' is omitted.
$W_{[s,t]}$ and $P_{r,s}$ represent the swap matrix and power-reducing matrix defined as
$W_{[s,t]}:=[I_t\otimes\delta_s^1~\cdots~I_t\otimes\delta_s^s]$ and
$P_{r,s}:=\hbox{diag}\{\delta_s^1,\cdots,\delta_s^s\}$,
respectively. For more properties on the swap matrix and power-reducing matrix, please refer
to \cite{cheng2011}.

\section{PROBLEM FORMULATION}

\begin{figure}[thpb]
      \centering
%      \framebox{\parbox{3in}{We suggest that you use a text box to insert a graphic (which is ideally a 300 dpi TIFF or EPS file, with all fonts embedded) because, in an document, this method is somewhat more stable than directly inserting a picture.
%}}
      \includegraphics[scale=0.5]{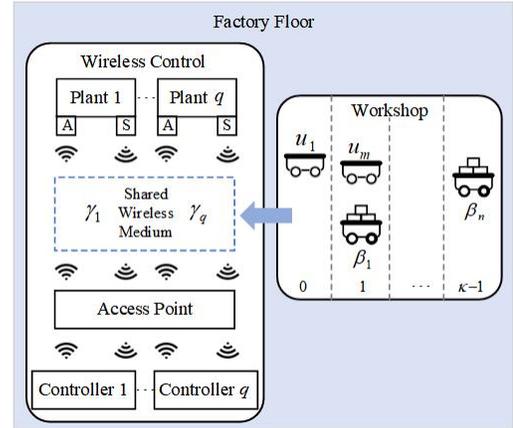}
\vskip-5mm
      \caption{Framework of a heterogeneous IIoT system where a WCS and an MAS
coordinate with each other to jointly perform overall task. `S' and `A' respectively
represent sensing and actuation capabilities. The local channel state of each system $i$ ($\gamma_i\in\mathcal{D}_s$) in WCS is
dependent on the positions of state agents
($\beta_j\in\mathcal{D}_\kappa$ represents the position of state agent $j$, $j=1,\cdots,n$) and positions of control agents
($u_l\in\mathcal{D}_\kappa$ represents the position of control agent $n+l$, $l=1,\cdots,m$).}
%, denotes the local channel
%state of system $i$ at time $k$, and $u_l(k)$, $\beta_j(k)\in\mathcal{D}_\kappa$ respectively
%represent the positions of control agent $l$ and state agent $j$ at time $k$, $l=1,\cdots,m$,
%$j=1,\cdots,n$.}
     \label{fig01}
   \end{figure}

In this paper, we consider a heterogeneous IIoT system where a WCS and an MAS
coordinate with each other to jointly perform overall task in a typical smart manufacturing
scenario (Fig. \ref{fig01}).
In the WCS, $q$ independent plants are controlled over a shared wireless medium, where the
measurements of plant $i$, $i\in\{1,\cdots,q\}$ are wirelessly transmitted to the access point by sensor $i$ to compute the
control inputs. At the same time, several mobile agents perform manufacturing tasks in the
nearby automation cell, i.e., the workshop, which is discretized into $\kappa$ two dimensional regions, denoted
by $0,1,\cdots,\kappa-1$. $m+n$ mobile agents are classified into two groups as
state agents $\{1,\cdots,n\}$ and control agents $\{n+1,\cdots,n+m\}$, where state agents are responsible for completing tasks such as inspection, monitoring, sensing and
transporting, while control agents are responsible for controlling the evolutionary behavior
of the MAS. In this process, the positions
of mobile agents may lead to shadow fading on the wireless channel that the WCS is controlled
over, resulting in the coupling between the WCS and MAS.
\subsection{Description of the WCS}

Denote the state of system $i$ at time $k$ by $x_i(k)\in\mathbb{R}^{n_i}$, $i=1,\cdots,q$.
Assume that the
evolution of each system $i$ depends on whether a
transmission occurs at time $k$ or not. Using $\lambda_{i}(k)\in\mathcal{D}_2$ to indicate
the successful or fail transmission of system $i$ at time $k$, then the dynamics of the WCS
is described as the following switched model:
\be\label{eq01}
x_i(k+1)=\left\{
  \begin{array}{ll}
    A_{c,i}x_i(k)+\xi_i(k), & \lambda_i(k)=1, \\
    A_{o,i}x_i(k)+\xi_i(k), & \lambda_i(k)=0,
  \end{array}
\right.~i=1,\cdots,q.
\ee
In (\ref{eq01}), the dynamics of each system $i$ at a successful transmission and a fail
transmission are characterized by matrices $A_{c,i}$ and
$A_{o,i}\in \mathbb{R}^{n_i\times n_i}$, respectively, where `c' and `o' respectively
represent the closed-loop and open-loop. The additive terms $\xi_i(k)$ represent
an independent identically distributed noise process with mean zero and covariance
$\Xi_i\in S_+^{n_i}$. The above WCS description (\ref{eq01})
can model a variety of control operations \cite{gatsis2015}. In this model, the closed-loop
dynamics for all $q$ systems
are fixed, that is, adequate controllers have been pre-designed.

At every slot $k$ each sensor $i$ chooses whether transmitting
over the shared channel or not adapting to its local channel state.
Discretize the wireless channel state of the WCS into $s$ values with each value representing
a channel state interval, denoted by $0,1,\cdots,s-1$,
and denote the local channel state of system $i$ at slot $k$ by
$\gamma_i(k)\in\mathcal{D}_s$.
Assume that the wireless communication policy is pre-designed,
and suppose that it holds
\be\label{eq04}
\alpha_i(k)=h_i(\gamma_i(k)),
\ee
where $h_i:\mathcal{D}_s\rightarrow\mathcal{D}_2$, $\alpha_i(k)=1$ (or $0$) means that
sensor $i$ transmits (or does not transmit) over the shared channel at slot $k$.

At slot $k$, if sensor $i$ chooses to transmit over the shared channel, assume that it selects a
transmit power level $\mu_i(k)\in\{\mu_{i,1},\cdots,\mu_{i,p_i}\}$ according to the
following rule:
\be\label{eq06}
\mathbb{P}\{\mu_i(k)=\mu_{i,a}|\alpha_i(k)=1,\gamma_i(k)=b\}=\mu_{a,b}^i,
\ee
where $a\in\{1,\cdots,p_i\}$, $b\in\mathcal{D}_s$, $\mu_{a,b}^i\in[0,1]$ and $\mathbb{P}$ is the probability.

The sensor's transmission might fail due to packet decoding errors and packet collisions.
We call a collision happens on system $i$, if some other sensor $j\neq i$ transmits in the
same time slot. %Thus, the probability that sensor $i$'s transmission is free of collisions, i.e., that no other sensor transmits and causes collisions on link $i$, equals $\prod_{j\neq i}(1-\mathbb{P}\{\alpha_j(k)=1\})$.
Use $\eta_i(k)\in\mathcal{D}_2$ to indicate the successful or fail packet decoding at the access point
of system $i$ at time $k$. Then, if sensor $i$ transmits at slot $k$, and no other sensor transmits, then the
relationship among the packet decoding success, local channel state and transmission power is
\be\label{eq08}
\mathbb{P}\{\eta_i(k)=1|\gamma_i(k)=a,\mu_i(k)=\mu_{i,b}\}=\eta_{a,b}^i,
\ee
where $a\in\mathcal{D}_s$ and $\eta_{a,b}^i\in[0,1]$.

\subsection{Description of the MAS}

%In this paper, we adopt the FFNs architecture to proceed the operational control for MASs.
%Discretize the workshop of the MAS into $\kappa$ two dimensional regions
%where $\kappa$ is a prime number.
%In addition, classify the agents into two groups as state agents and control agents,
%where state agents are responsible for completing tasks such as inspection, monitoring, sensing and
%transporting, while control agents are responsible for control the evolutionary behavior
%of the MAS,
%and denote their states at time $k$ by $\beta_i(k)$ and $u_l(k)$, respectively.
We adopt the FFN architecture \cite{Pasqualetti2014} to model the operational control for MASs
with limited communication, memory and computation capabilities.
For the state agents, the state is updated as a weighted combination of
its own value and those of its in-neighbors including state agents and control agents. More
specifically, it holds that
\begin{align}\nonumber
&\beta_i(k+1)=\{_\kappa\sum\limits_{j=1}^n\} a_{i,j}^{\sigma(k)}\times_{\kappa}\beta_j(k)\\\label{eq02}
&~~~~~~~+_{\kappa}\{_\kappa\sum\limits_{l=1}^m\} b_{i,l}^{\sigma(k)}\times_{\kappa}u_l(k), i=1,\cdots,n,
\end{align}
where $\kappa$ is a prime number, $a_{i,j}^{\sigma(k)}$, $b_{i,l}^{\sigma(k)}\in\mathcal{D}_\kappa$, $i,j=1,\cdots, n$,
$l=1,\cdots,m$, $\sigma: \mathbb{N}\rightarrow \{1,\cdots,w\}$ is the profile-dependent
switching signal, and operations ``$+_\kappa$'' and ``$\times_\kappa$'' are modular addition
and modular multiplication over $\mathcal{D}_\kappa$, respectively \cite{Lidl1996}.
$\beta(k):=(\beta_1(k),\cdots,\beta_n(k))\in C_\beta\subseteq\mathcal{D}_\kappa^n$ and
$u(k):=(u_1(k),\cdots,u_m(k))\in C_u(\beta(k))\subseteq\mathcal{D}_\kappa^m$ represent the
state profile and control profile, respectively, where $C_\beta$ and $C_u(\beta)$,
$\beta\in C_\beta$ respectively denote state constraints and control constraints.

 %and
%$z(k):=(u(k),\beta(k))\in\mathcal{D}_\kappa^{m+n} $ denotes the profile of FFN (\ref{eq02}).
%Set $C_u:=\{C_u(\beta): \beta\in C_\beta\}$. We call $C_\beta$ and $C_u$ the state
%constraint and control constraint, respectively.
Consider the state feedback controller of FFN (\ref{eq02}) in the following form:
\be\label{eq010}
u(k)=\pi(\beta(k)),
\ee
where $\pi:\mathcal{D}_\kappa^{n}\rightarrow \mathcal{D}_\kappa^m$. Denote the profile of
FFN (\ref{eq02}) by
$z(k):=(u(k),\beta(k))\in\mathcal{D}_\kappa^{m+n}$. Given an initial
state profile $\beta(0)=\beta_0\in C_\beta$ and a state feedback law $\pi$, FFN
(\ref{eq02}) becomes a closed-loop system and one can obtain $z(k;\beta_0,\pi)$,
$k\in\mathbb{N}$. If for any initial state profile $\beta_0\in C_\beta$,
$z(k;\beta_0,\pi)\in \{(u,\beta):\beta\in C_\beta, u\in C_u(\beta)\}$ holds for any
$k\in\mathbb{N}$, we call $\pi$ an admissible state feedback law. Denote the set of all
admissible state feedback laws $\pi$ by $\Pi$.

\subsection{WCS and MAS Coupling}

%As shown in Fig. \ref{fig01}, WCS (\ref{eq01}) and MAS (\ref{eq02}) interact with each other
%to jointly perform overall task. In this process, the coupling between the two systems lies
%in that the positions of mobile agents
%modeled by the state and control profiles of FFN (\ref{eq02}) may lead to shadow fading on the
%wireless channel that the WCS controlled over.

%In this process, the positions
%of mobile agents may lead to shadow fading on the wireless channel that the WCS controlled
%over, resulting in the coupling between the WCS and MAS.

When the WCS and the MAS coordinate with each other to jointly perform overall task,
the positions of mobile agents may lead to shadow fading on the wireless channel that the WCS
is controlled over.
%Given the dependence of local channel state for system $i$ on the positions of
%mobile agents as follows:
We model such state-dependent fading channel as follows:
\be\label{eq03}
\mathbb{P}\{\gamma_i(k)=a|z(k)=z\}=\gamma_{a,z}^{i},
\ee
where $z\in\mathcal{D}_\kappa^{n+m}$ and $\gamma_{a,z}^i\in[0,1]$.
According to (\ref{eq04})-(\ref{eq08}),
%the local channel state influences whether or not the
%sensor transmits over the shared channel and the probability of packet decoding success, and
%finally influences the probability of successful wireless transmission. Thus,
the positions
of the mobile agents influence the probability of successful wireless transmission for the
WCS. The WCS and MAS coupling can be expressed as
%and the relationship can be expressed as
 %Thus, the WCS and MAS coupling can be expressed as
%Then, according to (\ref{eq04})-(\ref{eq08}) and (\ref{eq03}),  the
%WCS and MAS coupling
%can be expressed as
\begin{align}\nonumber
&\mathbb{P}\{\lambda_i(k)=1|z(k)=z\}\\\nonumber
=&\mathbb{P}\{\alpha_i(k)=1|z(k)=z\}\mathbb{P}\{\eta_i(k)=1|z(k)=z\}\\\nonumber
&\times\prod_{j\neq i}(1-\mathbb{P}\{\alpha_j(k)=1|z(k)=z\})\\\label{eq09}
=&\alpha_{z}^i\eta_{z}^i\prod_{j\neq i}(1-\alpha_{z}^j):=\lambda_z^i,
\end{align}
where $\alpha_{z}^i:=\sum_{a=0}^{s-1}\gamma_{a,z}^{i}h_i(a)$, $\bar{\mu}_{b,z}^i:=\sum_{c=0}^{s-1}\gamma_{c,z}^{i}\mu_{b,c}^i$
and $\eta_z^i:=\sum_{(a,b)\in \mathcal{D}_s\times\{1,\cdots,p_i\}}\gamma_{a,z}^{i}\bar{\mu}_{b,z}^i\eta_{a,b}^i$.
%\begin{eqnarray}\nonumber
%\alpha_{z}^i:&=&\mathbb{P}\{\alpha_i(k)=1|z(k)=z\}=\sum_{a=1}^s\gamma_{a,z}^{i}h_i(a),\\\nonumber
%\eta_z^i:&=&\mathbb{P}\{\eta_i(k)=1|z(k)=z\}=\sum_{(a,b)\in D}\gamma_{a,z}^{i}\bar{\mu}_{b,z}^i\eta_{a,b}^i,\\\nonumber
%\bar{\mu}_{b,z}^i:&=&\mathbb{P}\{\mu_i(k)=\mu_{i,b}|z(k)=z\}=\sum_{c=1}^s\gamma_{c,z}^{i}\mu_{b,c}^i
%\end{eqnarray}
%and $D:=\mathcal{D}_s\times\langle1,p_i\rangle$.

%By taking into account of the WCS and MAS coupling,
Eq. (\ref{eq09}) presents
a state-dependent channel model. The probability of a successful transmission for
WCS (\ref{eq01}) depends on the profile of FFN (\ref{eq02}). It
is quite different from the traditional shadow fading model
\cite{Gatsis2014,zhangq1999}
%which
%invalidates the traditional methods on performance analysis of WCS (\ref{eq01})
assuming the channel states independent from the physical states.
Such coupling makes the traditional methods on performance analysis of WCS
\cite{gatsis2015,gatsis2018,Hristu2001,Molin,zhang2006,le2011} fail in the scenario considered
in this paper.

\subsection{Problem Formulation}
%In this paper, we aim to design controllers for FFN (\ref{eq02}) to ensure the stability for WNCS (\ref{eq01}).
Our primary goal in this paper is to design admissible controllers for MAS (\ref{eq02})
to satisfy certain level of closed-control performance of the WCS (\ref{eq01}) with respect
to (\ref{eq02}) in the presence of coupling (\ref{eq09}).
To this end, we introduce the Lyapunov function $V_i(x_i)=x_i^\top Q_ix_i$,
$Q_i\in S_{++}^{n_i}$ and consider the following Lyapunov-like
performance requirements for WCS (\ref{eq01}).

{\definition\label{def04} Given $Q=\{Q_1,\cdots,Q_q\}$ and $\rho=\{\rho_1,\cdots$, $\rho_q\}$ with
$\rho_i\in(0,1)$, $i=1,\cdots,q$, WCS (\ref{eq01}) is said to be $(Q,\rho)$-asymptotically
stable with respect to FFN (\ref{eq02}), if there exist a state feedback law $\pi\in \Pi$
and a positive integer $T$ such that the inequalities
\begin{align}\nonumber
&\mathbb{E}[V_i(x_i(k+1))|x_i(k),z(k;\beta_0,\pi)]\leq\rho_iV_i(x_i(k))\\\label{eq014}
&~~~~~~~~~~~~~~~~~~+Tr(Q_i\Xi_i), \forall\ x_i(k)\in\mathbb{R}^{n_i}
\end{align}
hold for any initial state profile $\beta_0\in C_\beta$, any $i\in\{1,\cdots,q\}$ and any
$k\geq T$, where $\mathbb{E}$ is the expectation. In (\ref{eq014}), $\rho_i$ represents desired decay rate and $Tr(Q_i\Xi_i)$ denotes the trace
of matrix $Q_i\Xi_i$ representing a persistent noise perturbation.}
%Moreover, we call $\pi$ satisfying (\ref{eq014}) a $(Q,\rho)$-state feedback
%law.}

%Given $Q$ and $\rho$, assume that WCS (\ref{eq01}) is $(Q,\rho)$-asymptotically stable
%without regard to the cyber-physical coupling with FFN (\ref{eq02}).
%%In fact, the cyber-physical coupling may degrade the control performance of WNCS (\ref{eq01}) due to shadow fading. Thus,
%The goal of this paper is to propose a criterion on the $(Q,\rho)$-asymptotically
%stablility of WCS (\ref{eq01}) with respect to FFN (\ref{eq02}), which provides reference
%for the heterogeneous IIoT system design. Moreover, if WCS (\ref{eq01}) is
%$(Q,\rho)$-asymptotically stable with respect to FFN (\ref{eq02}), we aim to present the
%method for constructing $(Q,\rho)$-state feedback laws.

\section{MAIN RESULTS}
In this section, based on ASSR approach and graph theory, we design state feedback
controllers for the MAS to ensure the Lyapunov-like performance of WCS in the presence of
WCS and MAS coupling.

\subsection{ASSR Reformulation of the MAS}
%In FFNs, the state of each agent takes values from a finite field $\mathcal{D}_\kappa$. In
%addition, the state of each state agent is updated as a
%weighted combination over $\mathcal{D}_\kappa$ of its own value and those of its
%in-neighbors with operations being performed modulo $\kappa$.
Due to the variables taking values from finite fields and the operations being
modular addition and modular multiplication, it is difficult to apply classical nonlinear system theory to
investigate FFNs.
The ASSR approach is a powerful tool which builds a bridge between finite-value systems and
classical control theory \cite{cheng2011}.
%By virtue of ASSR approach, numerous excellent results have been proposed for
It has been applied to Boolean
networks and logical control networks, finite games,
finite automata and so on \cite{li2018}, \cite{Liy2019}.
%Especially, the ASSR approach is also
%effective in studying FFNs  \cite{Liy2019}.
In this part, by virtue of ASSR method, we convert FFN (\ref{eq02}) into its
equivalent algebraic form, which facilitates the sequential studies.

Represent $i-1\in\mathcal{D}_\kappa$ by a vector $\delta_\kappa^i$, $i=1,\cdots,\kappa$.
%identify $i-1\sim\delta_\kappa^i$.
In addition, represent the switching signal $\sigma(k)=i$ by $\delta_w^i$, $i=1,\cdots,w$.
Then, $\beta_i(k),u_i(k)\in\mathcal{D}_\kappa$ and $\sigma(k)\in\{1,\cdots,w\}$ have their corresponding
vector forms (still using the same symbols) $\beta_i(k),u_i(k)\in\Delta_\kappa$ and
$\sigma(k)\in \Delta_w$,
respectively. Letting $\beta(k)=\ltimes_{i=1}^n\beta_i(k)$, $u(k)=\ltimes_{i=1}^m u_i(k)$
and $z(k)=u(k)\beta(k)$, we have $C_\beta\subseteq\Delta_N$, $C_u(\beta(k))\subseteq\Delta_M$,
where $N:=\kappa^n$ and $M:=\kappa^m$.
In addition, the state feedback law $\pi$ has the form $\pi: \Delta_N\rightarrow \Delta_M$.

The following lemma presents the structural matrices for ``$+_\kappa$'' and
``$\times_\kappa$'' \cite{Liy2019}.
{\lemma\label{lemma01} Let $a,b\in\mathcal{D}_\kappa\sim\Delta_\kappa$.
\begin{enumerate}
\item[\rm(i)] $a+_\kappa b=F_{+_\kappa}\ltimes a\ltimes b$, where $F_{+_\kappa}=\delta_\kappa[A_1~A_2~\cdots~A_\kappa]$,
$A_1=(1~\cdots~\kappa)$ and $A_j=(j~\cdots~\kappa~1~\cdots~j-1)$, $j=2,\cdots,\kappa$.
\item[\rm(ii)] $a\times_\kappa b=F_{\times_\kappa}\ltimes a\ltimes b$, where
$F_{\times_\kappa}=\delta_\kappa[B_1~B_2~\cdots~B_\kappa]$,
$B_j=((0\times(j-1))\hbox{mod}(\kappa)+1~~(1\times(j-1))\hbox{mod}(\kappa)+1$ $\cdots~~((\kappa-1)\times(j-1))\hbox{mod}(\kappa)+1)$, $j=1,\cdots,\kappa$.
\end{enumerate}}

Define
\be\label{eq012}
A_{i,j}=[a_{i,j}^1~\cdots~a_{i,j}^w],~B_{i,l}=[b_{i,l}^1~\cdots~b_{i,l}^w],
\ee
$i,j=1,\cdots,n$, $l=1,\cdots,m$. Then, using the vector form of elements in
$\mathcal{D}_\kappa$ and $\{1,\cdots,w\}$, for the dynamics of the $i$-th agent in FFN (\ref{eq02}),
we have
\begin{align}\nonumber
\beta_i(k+1)=&\{_\kappa\sum\limits_{j=1}^n\} a_{i,j}^{\sigma(k)}\times_{\kappa}\beta_j(k)+_{\kappa}\{_\kappa\sum\limits_{l=1}^m\} b_{i,l}^{\sigma(k)}\times_{\kappa}u_l(k)\\\nonumber
=&F_{+_\kappa}^n\ltimes_{j=1}^n(F_{\times_\kappa}A_{i,j}\sigma(k)\beta_j(k))F_{+_\kappa}^{m-1}\\\label{eq026}
&\ltimes_{j=1}^m(F_{\times_\kappa}B_{i,j}\sigma(k)u_j(k)).
\end{align}
Skipping simple derivations, one can obtain
\begin{align}\nonumber
&\ltimes_{j=1}^n(F_{\times_\kappa}A_{i,j}\sigma(k)\beta_j(k))\\\nonumber
=&F_{\times_\kappa}A_{i,1}\ltimes_{j=1}^{n-1}\Big\{[I_{w\kappa^j}\otimes(F_{\times_\kappa}A_{i,j+1})]W_{[w,w\kappa^j]}P_{r,w}\Big\}\\\nonumber
&\ltimes\sigma(k)\beta(k)\\\nonumber
:=&F_{i,1}\sigma(k)\beta(k).
\end{align}
Similarly, we have
\begin{align}\nonumber
&\ltimes_{j=1}^m(F_{\times_\kappa}B_{i,j}\sigma(k)u_j(k))\\\nonumber
=&F_{\times_\kappa}B_{i,1}\ltimes_{j=1}^{m-1}\Big\{[I_{w\kappa^j}\otimes(F_{\times_\kappa}B_{i,j+1})]W_{[w,w\kappa^j]}P_{r,w}\Big\}\\\nonumber
&\ltimes\sigma(k)u(k)\\\nonumber
:=&F_{i,2}\sigma(k)u(k).
\end{align}
Then, for (\ref{eq026}), it holds
\begin{align}\nonumber
\beta_i(k+1)=&F_{+_\kappa}^nF_{i,1}\sigma(k)\beta(k)F_{+_\kappa}^{m-1}F_{i,2}\sigma(k)u(k)\\\nonumber
=&F_{+_\kappa}^nF_{i,1}[I_{wN}\otimes(F_{+_\kappa}^{m-1}F_{i,2})]\sigma(k)\beta(k)\sigma(k)u(k)\\\nonumber
=&F_{+_\kappa}^nF_{i,1}[I_{MN}\otimes(F_{+_\kappa}^{m-1}F_{i,2})](I_w\otimes W_{[wM,N]})\\\label{eq027}
&\ltimes P_{r,w}\sigma(k)z(k):=F_{i,3}\sigma(k)z(k).
\end{align}

In this paper, we consider the case where the switching signal $\sigma(k)$ is profile
dependent, namely, $\sigma(k)$ is determined by the
profile $z(k)$. Then, the switching signal has the following form:
\be\nonumber
\sigma(k)=\Theta z(k),
\ee
which together with (\ref{eq027}) shows
\be\nonumber
\beta_i(k+1)=F_{i,3}\Theta P_{r,MN}z(k):=F_iz(k),
\ee
where $\Theta\in\mathcal{L}_{w\times MN}$ and $F_i\in\mathcal{L}_{\kappa\times MN}$.
Multiplying the dynamics of $n$ state agents in FFN (\ref{eq02}), one can obtain the
following equivalent algebraic form of FFN (\ref{eq02}):
\be\label{eq015}
\beta(k)=Fz(k),
\ee
where $F\in\mathcal{L}_{N\times MN}$ satisfying $Col_l(F)=\ltimes_{i=1}^nCol_l(F_i)$,
$l=1,\cdots,MN$. Split matrix $F$ into $M$ equal blocks as
$$F=[Blk_1(F)~Blk_2(F)~\cdots~Blk_M(F)].$$
Then, $[Blk_l(F)]_{a,b}=1$ means that under the
control profile $\delta_M^l$, state profile $\delta_N^b$ can reach state profile
$\delta_N^a$ in one step.

Similarly, we can obtain the following equivalent algebraic form for state feedback
controller (\ref{eq010}):
\be\label{eq016}
u(k)=L\beta(k),
\ee
where $L\in\mathcal{L}_{M\times N}$ is the state feedback gain matrix.

In addition, using the vector form of $z(k)$, the WCS and MAS coupling (\ref{eq09})
can be converted into an equivalent form
\be\label{eq09-1}
\mathbb{P}\{\lambda_i(k)=1|z(k)\}=\Lambda_iz(k),
\ee
where $\Lambda_i=[\lambda_{z_1}^{i}~\lambda_{z_2}^{i}~\cdots~\lambda_{z_{MN}}^{i}]$
with $\lambda_{z_j}^{i}$ given in (\ref{eq09}),  %$\Lambda_i^\top\in\mathbb{R}^{MN}$ with $Col_j(\Lambda_i):=\lambda_{z_j}^{i}$,
and $z_j$ has vector form $\delta_{MN}^j$, $j\in\{1,\cdots,MN\}$.

\subsection{Lyapunov-like Performance Analysis of the WCS}
In this part, in order to facilitate the controller design for the MAS, we convert
Lyapunov-like
performance requirements (\ref{eq014}) of WCS (\ref{eq01}) to
a specific control objective of FFN (\ref{eq015}).

Consider Lyapunov-like
performance requirements (\ref{eq014}). Since $\mathbb{E}[\xi_i(k)]=\mathbf{0}$
and $\lambda_i(k)$ is independent of
$x_i(k)$, it holds
\begin{align}\nonumber
&\mathbb{E}[V_i(x_i(k+1))|x_i(k),z(k;\beta_0,\pi)]\\\nonumber
=&\mathbb{P}\{\lambda_i(k)=1|z(k;\beta_0,\pi)\}x_i^\top(k)A_{c,i}^\top Q_i A_{c,i}x_i(k)\\\nonumber
&+\mathbb{P}\{\lambda_i(k)=0|z(k;\beta_0,\pi)\}x_i^\top(k)A_{o,i}^\top Q_i A_{o,i}x_i(k)\\\nonumber
&+T_r(Q_i\Xi_i).
\end{align}
Then, condition (\ref{eq014}) is equivalent to the condition that for all $x_i(k)\neq \mathbf{0}$,
\begin{align}\nonumber
&\mathbb{P}\{\lambda_i(k)=1|z(k;\beta_0,\pi)\}\\\nonumber
\geq&\frac{x_i^\top(k)(A_{o,i}^\top Q_i A_{o,i}-\rho_iQ_i)x_i(k)}{x_i^\top(k)(A_{o,i}^\top Q_iA_{o,i}-A_{c,i}^\top Q_iA_{c,i})x_i(k)},
\end{align}
which is equivalent to
\be\label{eq021}
\mathbb{P}\{\lambda_i(k)=1|z(k;\beta_0,\pi)\}\geq s_i,
\ee
where
\be\label{eq053}
s_i=\sup\limits_{y\in\mathbb{R}^{n_i},y\neq 0}\frac{y^\top(A_{o,i}^\top Q_iA_{o,i}-\rho_iQ_i)y}{y^\top(A_{o,i}^\top Q_iA_{o,i}-A_{c,i}^\top Q_iA_{c,i})y}.
\ee
It is noted that $s_i$ represents the lower bound of the probability of successful
transmission for system $i$ that ensures the desired Lyapunov decay rate $\rho_i$ under the
WCS and MAS coupling.

Based on the above representation, we can then define $\Omega(s)$ consisting of all
profiles of FFN (\ref{eq015}) in $C_z$ under the influence of which the probability of
successful transmission for each system $i$ is no less than $s_i$, that is,
\be\label{eq050}
\Omega(s):=\cap_{i=1}^q\{\delta_{MN}^j: Col_j(\Lambda_i)\geq s_i\}\cap C_z,
\ee
where $s=[s_1~\cdots~s_q]$, $C_z:=\{z=u\beta\in\Delta_{MN}:\beta\in C_\beta,u\in C_u(\beta)\}$ and $\Lambda_i$
is defined in (\ref{eq09-1}). It follows from (\ref{eq09-1}) and (\ref{eq021}) that $\Omega(s)$ consists of
all profiles of FFN (\ref{eq015}) ensuring the desired Lyapunov decay rates for all
systems in WCS (\ref{eq01}).

%Define
%\be\label{eq050-1}
%\Omega(s_i):=\{\delta_{MN}^j: Col_j(\Lambda_i)\geq s_i\}\cap C_z
%\ee
%and
%\be\label{eq050}
%\Omega(s):=\cap_{i=1}^q\Omega(s_i),
%\ee
%where $s=[s_1~\cdots~s_q]$, $C_z:=\{z=u\beta\in\Delta_{MN}:\beta\in C_\beta,u\in C_u(\beta)\}$ and $\Lambda_i$
%is defined in (\ref{eq09-1}).

%According to (\ref{eq09-1}) and (\ref{eq021}), it is obvious that $\Omega(s)$ consists of
%all profiles of FFN (\ref{eq015}) ensuring the desired Lyapunov decay rates for all systems
%in WCS (\ref{eq01}).
%Then, by Definition \ref{def04}, WCS
%(\ref{eq01}) is $(Q,\rho)$-asymptotically stable with respect to
%FFN (\ref{eq015}), if and only if there exists a state feedback law $\pi\in \Pi$
%and a positive integer $T$ such that the profile $z(k;\beta_0,\pi)\in\Omega(s)$
%holds for any initial state profile $\beta_0$ and any $k\geq T$.
%In this case, we call FFN (\ref{eq015}) achieves constrained control-state
%$\Omega(s)$-stabilizable.

{\definition\label{def01} Given $\mathcal{M}\subseteq\Delta_{MN}$, FFN (\ref{eq015}) is
called constrained control-state $\mathcal{M}$-stabilizable, if there exist a state
feedback law $\pi\in\Pi$ and a positive integer $T$ such that
$z(k;\beta_0,\pi)\in\mathcal{M}$ holds for any integer $k\geq T$ and any initial state
profile $\beta_0\in C_\beta$.}

We have the following result on the $(Q,\rho)$-asymptotically stability of
WCS (\ref{eq01}).

{\lemma\label{lemma020} WCS (\ref{eq01}) is $(Q,\rho)$-asymptotically stable with respect to
FFN (\ref{eq015}), if and only if FFN (\ref{eq015}) is constrained control-state
$\Omega(s)$-stabilizable, where $\Omega(s)$ is defined in (\ref{eq050}).}
\proof\ The result can be directly obtained from (\ref{eq09-1}), (\ref{eq021}) and the
construction of $\Omega(s)$, so we omit the proof here.
$\Box$

Note that the constrained control-state stabilization imposes a coupled requirement
of control profile and state profile on controller design.
A natural idea is to adopt the constrained set stabilization, which requires the state
profiles converging to a given set. It is more convenient for controller design than constrained control-state set
stabilization.
%For the definition of constrained control-state set stabilization, if we set
%$\mathcal{M}\subseteq\Delta_N $ and replace $z(k;\beta_0,\pi)\in\mathcal{M}$ by
%$\beta(k;\beta_0,\pi)\in\mathcal{M}$, then it becomes the definition of
%constrained set stabilization.
%Since constrained set stabilization only requires the state profiles converging to a given
%set, it is more convenient for controller design than constrained control-state set
%stabilization. Thus, a natural idea is to convert the constrained control-state set
%stabilization requirement in Lemma \ref{lemma020} to constrained set stabilization.
To this end, we introduce the definition of constrained control invariant set (CCIS).

%Note that in constrained control-state set stabilization, the control objective is proposed
%based on profile $z(k)$, which is inconvenient for controller design. The concept of
%constrained control-state set stabilization is similar to constrained control-state
%set stabilization. Given $\mathcal{M}\subseteq\Delta_N$, FFN (\ref{eq015})
%is called constrained $\mathcal{M}$-stabilizable, if there exist a state feedback law
%$\pi\in\Pi$ and a positive integer $T$ such that $\beta(k;\beta_0,\pi)\in\mathcal{M}$
%holds for any integer $k\geq T$ and any initial state profile $\beta_0\in C_\beta$.
%The set stabilization of finite-value systems have been extensively studied
%\cite{guo2015,zhangx2020,gao2021,lif2017-1,zhangq2021,liu2020}. Different from the constrained control-state set
%stabilization, the constrained set
%stabilization of FFN (\ref{eq015}) only requires that the state profiles converge to a given
%set. Thus, a natural idea is to convert the constrained control-state set stabilization to
%constrained set stabilization. To this end, we introduce the definition of constrained
%control invariant set (CCIS).

For $\mathcal{M}\subseteq\Delta_{MN}$, define
\begin{align}\nonumber
&\Phi(\mathcal{M}):=\{\beta\in C_\beta: \hbox{there exists}\ u\in C_u(\beta)\ \hbox{such that}\\\label{eq020}
&~~~~~~~~~~~~~~u\beta\in\mathcal{M}\}.
\end{align}
Correspondingly, for any $\beta\in\Phi(\mathcal{M})$, define
\be\label{eq017}
C_u^{\mathcal{M}}(\beta)=\{u\in C_u(\beta):u\beta\in\mathcal{M}\}.
\ee

{\definition\label{def03} Given $\mathcal{M}\subseteq\Delta_{MN}$, a subset
$\mathcal{C}\subseteq \Phi(\mathcal{M})$ is called a CCIS
of FFN (\ref{eq015}) with respect to $\mathcal{M}$, if for any $\beta_0\in \mathcal{C}$,
there exists a control sequence $u=\{u(k)\in C_u^{\mathcal{M}}(\beta(k)): k\in\mathbb{N}\}$ such that
$\beta(k;\beta_0,u)\in\mathcal{C}$ holds for any $k\in\mathbb{N}$.}

Obviously, the union of any two CCISs of FFN (\ref{eq015}) with respect to $\mathcal{M}$
is another CCIS with respect to $\mathcal{M}$. The union of all CCISs with respect to
$\mathcal{M}$ is called its largest constrained control invariant
set (LCCIS) with respect to $\mathcal{M}$, denoted
by $I(\mathcal{M})$.

By virtue of LCCIS, the following result establishes the equivalence between constrained
control-state set stabilization and constrained set stabilization.

{\lemma\label{lemma010} FFN (\ref{eq015}) is constrained control-state
$\mathcal{M}$-stabilizable, if and only if it is constrained $I(\mathcal{M})$-stabilizable.}
\proof\
(Necessity) Since $I(\mathcal{M})$ is the LCCIS with respect to $\mathcal{M}$, for any
$\beta_0\in\Phi(\mathcal{M})\setminus I(\mathcal{M})$ and any control sequence
$u=\{u(k)\in C_u^{\mathcal{M}}(\beta(k)): k\in\mathbb{N}\}$, there exists an integer
$k(\beta_0,u)$ such that
\be\label{eq019}
\beta(k(\beta_0,u);\beta_0,u)\notin\Phi(\mathcal{M}).
\ee

Next, we prove the conclusion by absurdity.

If FFN (\ref{eq015}) is constrained control-state $\mathcal{M}$-stabilizable, then
we can obtain a feasible state feedback law $\pi$ and a positive integer $T$.
%there exist a state feedback law $\pi\in\Pi$ and a positive integer $T$ such that
%$z(k;\beta_0,\pi)\in\mathcal{M}$ holds for any integer $k\geq T$ and any initial state
%profile $\beta_0\in C_\beta$.
Then, for any integer $k\geq T$ and any initial state
profile $\beta_0\in C_\beta$, it holds that
\be\label{eq018}
\beta(k;\beta_0,\pi)\in\Phi(\mathcal{M}),~\pi(\beta(k;\beta_0,\pi))\in C_u^{\mathcal{M}}(\beta(k;\beta_0,\pi)).
\ee
If there exist an integer $k_0\geq T$ and a state profile $\beta_0\in C_\beta$ satisfying
$\beta(k_0;\beta_0,\pi)\notin I(\mathcal{M})$, then according to (\ref{eq019}), setting
$u=\{\pi(\beta(k_0+k;\beta_0,\pi)): k\in\mathbb{N}\}$,
%for any $u\in\bar{C}_u(\beta(k;\beta_0,\pi))$,
it holds that $$\beta(k_0+k(\beta(k_0;\beta_0,\pi),u);\beta_0,\pi)\notin\Phi(\mathcal{M}),$$
which contradicts to (\ref{eq018}). Thus, for any integer $k_0\geq T$ and any initial state
profile $\beta_0\in C_\beta$, it holds that $\beta(k;\beta_0,\pi)\in I(\mathcal{M})$,
which together with $\pi\in\Pi$ shows that FFN (\ref{eq015}) is constrained
$I(\mathcal{M})$-stabilizable.

(Sufficiency) If FFN (\ref{eq015}) is constrained $I(\mathcal{M})$-stabilizable, then
there exist a state feedback law $\pi\in\Pi$ and a positive integer $T$ such that
$\beta(k;\beta_0,\pi)\in I(\mathcal{M})$ holds for any integer $k\geq T$ and any initial
state profile $\beta_0\in C_\beta$. Defining
\be\label{eq011}
\bar{\pi}(\beta):=\left\{
              \begin{array}{ll}
                 u\in C_u^{\mathcal{M}}(\beta),&  \beta\in I(\mathcal{M});\\
                \pi(\beta), &  \hbox{otherwise,}
              \end{array}
            \right.
\ee
it is obvious that $\bar{\pi}\in\Pi$. In addition, it is easy to obtain from (\ref{eq017})
that $z(k;\beta_0,\bar{\pi})\in \mathcal{M}$ holds for any integer $k\geq T$ and any
initial state profile $\beta_0\in C_\beta$. Thus, FFN (\ref{eq015}) is constrained
control-state $\mathcal{M}$-stabilizable.
$\Box$

Based on Lemmas \ref{lemma020} and \ref{lemma010}, the following result deduces the
$(Q,\rho)$-asymptotically stability of WCS (\ref{eq01}) to the constrained set stabilization
of FFN (\ref{eq015}).

{\theorem\label{th01} WCS (\ref{eq01}) is $(Q,\rho)$-asymptotically stable with respect
to FFN (\ref{eq015}) if and only if FFN (\ref{eq015}) is constrained
$I(\Omega(s))$-stabilizable, where $\Omega(s)$ is defined in (\ref{eq050}).}

\proof\ The result can be directly obtained from Lemmas \ref{lemma020} and \ref{lemma010},
so we omit the proof here.
%Since $\mathbb{E}[\xi_i(k)]=\mathbf{0}$ and $\lambda_i(k)$ is independent of
%$x_i(k)$, it holds that
%\begin{eqnarray}\nonumber
%&&\mathbb{E}[V_i(x_i(k+1))|x_i(k),z(k;\beta_0,\pi^\ast)]\\\nonumber
%=&&\mathbb{P}\{\lambda_i(k)=1|z(k;\beta_0,\pi^\ast)\}x_i^\top(k)A_{c,i}^\top Q_i A_{c,i}x_i(k)\\\nonumber
%&&+\mathbb{P}\{\lambda_i(k)=0|z(k;\beta_0,\pi^\ast)\}x_i^\top(k)A_{o,i}^\top Q_i A_{o,i}x_i(k)\\\nonumber
%&&+T_r(Q_i\Xi_i).
%\end{eqnarray}
%Then, condition (\ref{eq014}) is equivalent to
%\begin{eqnarray}\nonumber
%&&\mathbb{P}\{\lambda_i(k)=1|z(k;\beta_0,\pi^\ast)\}\\\nonumber
%&\geq&\frac{x_i^\top(k)(A_{o,i}^\top Q_i A_{o,i}-\rho_iQ_i)x_i(k)}{x_i^\top(k)(A_{o,i}^\top Q_iA_{o,i}-A_{c,i}^\top Q_iA_{c,i})x_i(k)},~\forall\ x_i(k)\neq \mathbf{0},
%\end{eqnarray}
%which is equivalent to
%\be\label{eq021}
%\mathbb{P}\{\lambda_i(k)=1|z(k;\beta_0,\pi^\ast)\}\geq s_i,
%\ee
%where
%$$s_i=\sup\limits_{y\in\mathbb{R}^{n_i},y\neq 0}\frac{y^\top(A_{o,i}^\top Q_iA_{o,i}-\rho_iQ_i)y}{y^\top(A_{o,i}^\top Q_iA_{o,i}-A_{c,i}^\top Q_iA_{c,i})y}.$$
%Thus, combing (\ref{eq021}) and (\ref{eq09-1}), WNCS (\ref{eq01}) is
%$(Q,\rho)$-asymptotically stable with respect to FFN (\ref{eq015}), if and only if there
%exist a state feedback law $\pi\in \Pi$ and a positive integer $T$, such that
%$z(k;\beta_0,\pi)\in\Omega(s)$ holds for any integer $k\geq T$ and any initial state
%profile $\beta_0$, that is, FFN (\ref{eq015}) is constrained control-state
%$\Omega(s)$-stabilizable. Therefore, from Lemma \ref{lemma010}, the conclusion follows.
$\Box$

%{\remark\label{rem01} Recently, based on the analysis of largest control invariant set
%(LCIS) contained in a given set $\mathcal{N}\subseteq\Delta_N$, denoted by
%$\tilde{I}(\mathcal{N})$, several novel criteria have been proposed for the set
%stabilization related problems of finite-value systems \cite{guo2015,gao2021,zhangx2020,liu2020}.
%Ignore the state and control constraints, that is, $C_\beta:=\Delta_N$,
%$C_u(\beta):=\Delta_M$, $\beta\in\Delta_N$, and denote the corresponding LCCIS with respect
%to $\mathcal{M}\subseteq\Delta_{MN}$ by $\bar{I}(\mathcal{M})$. Then,
%the difference
%between $\tilde{I}(\mathcal{N})$ and $\bar{I}(\mathcal{M})$ lies in that
%$\tilde{I}(\mathcal{N})$ is the largest set of state profiles that can remain in this set
%under a control $u\in\Delta_N$, while $\bar{I}(\mathcal{M})$ is the largest set of state
%profiles $\beta$ that can remain in this set under a control $u\in C_u^{\mathcal{M}}(\beta)$,
%that is,
%$\bar{I}(\mathcal{M})$ is an LCIS contained in $\Phi(\mathcal{M})$ with control constraints
%$C_u^{\mathcal{M}}(\beta)$, $\beta\in\Phi(\mathcal{M})$.
%%$C_u^{\mathcal{M}}:=\{C_u^{\mathcal{M}}(\beta): \beta\in\Phi(\mathcal{M})\}$.
%Thus,
%it is easy to see that $\bar{I}(\mathcal{M})\subseteq\tilde{I}(\Phi(\mathcal{M}))$.}

\subsection{A Graphical Method for Feasible Controller Design}

%In this part, we proceed the stability analysis for WNCS (\ref{eq01}) taking into account
%of cyber-physical coupling.%dependence of WNCS's stability on the state profile for MAS.

%\textbf{Graph-based approach is shown to be more computationally
%economical than algebraic methods on set stabilization problem \cite{gao2021}.}

In this part, we present a graph-based criterion on $(Q,\rho)$-asymptotically stability of
WCS (\ref{eq01}), which is much easier to verify.
%explore the verification of constrained set stabilization of
%FFN (\ref{eq015}) from the perspective of graph theory.
To this end, we introduce the
constrained state transition graph (STG) $G[C_\beta^0,C_u^0]=(V(G[C_\beta^0,C_u^0])$,
$E(G[C_\beta^0,C_u^0]))$ with respect to state constraints $C_\beta^0$ and
control constraints $C_u^0(\beta)$, $\beta\in C_\beta^0$.

%Given state constraints $C_\beta^0\subseteq C_\beta$ and
%control constraints $C_u^0(\beta)\subseteq \Delta_M$, $\beta\in C_\beta^0$.

%Represent each state profile in $C_\beta^0$ by a vertex and collect all vertices
%to form the vertex set $V(G[C_\beta^0,C_u^0])$. In addition, represent each constrained
%one-step state profile transition from $\beta_a\in C_\beta^0$ to $\beta_b\in C_\beta^0$ by
%a directed edge, denoted by $(\beta_a,\beta_b)$, and collect all directed edges to form a
%edge set $E(G[C_\beta^0,C_u^0])$, then
%\be\nonumber
%E(G[C_\beta^0,C_u^0])=\{(\beta_a,\beta_b)\in C_\beta^0\times C_\beta^0: \beta_b\in\mathcal{R}_1(\beta_a;C_u^0)\},
%\ee
%where $\mathcal{R}_1(\beta_a;C_u^0)$ denotes the one-step reachable set of state profile
%$\beta_a$ with respect to control constraint $C_u^0(\beta_a)$.
%In this part, let $\beta_a:=\delta_N^{a}$.
%Then, it holds that
%\begin{align}\nonumber
%\mathcal{R}_1(\beta_a;C_u^0):=&\{Fu\delta_N^{a}: u\in C_u^0(\delta_N^{a})\}\\\nonumber
%=&\{Col_{a}(Blk_l(F)):\delta_M^l\in C_u^0(\delta_N^{a})\}.
%\end{align}

Represent each state profile $\delta_N^{a}\in C_\beta^0$ by a vertex $v_a$ and collect all vertices
to form the vertex set $V(G[C_\beta^0,C_u^0])$. In addition, represent each constrained
one-step state profile transition from $\delta_N^{a}\in C_\beta^0$ to $\delta_N^{b}\in C_\beta^0$ by
a directed edge, denoted by $(v_{a},v_{b})$, and collect all directed edges to form a
edge set $E(G[C_\beta^0,C_u^0])$, then
\be\nonumber
E(G[C_\beta^0,C_u^0])=\{(v_{a},v_{b})\in C_\beta^0\times C_\beta^0: \delta_N^{b}\in\mathcal{R}_1(\delta_N^{a};C_u^0)\},
\ee
where $\mathcal{R}_1(\delta_N^{a};C_u^0)$ denotes the one-step reachable set of state profile
$\delta_N^{a}$ with respect to control constraint $C_u^0(\delta_N^{a})$.
Then, it holds that
\begin{align}\nonumber
\mathcal{R}_1(\delta_N^{a};C_u^0):=&\{Fu\delta_N^{a}: u\in C_u^0(\delta_N^{a})\}\\\nonumber
=&\{Col_{a}(Blk_l(F)):\delta_M^l\in C_u^0(\delta_N^{a})\}.
\end{align}

%For a directed graph $G$, denote its vertex set and edge set by $V(G)$ and $E(G)$,
%respectively. In addition, denote the directed edge from $v_i\in V(G)$ to $v_j\in V(G)$
%by an ordered pair $(v_i, v_j)$. Define $E^\top(G)=\{(v_j,v_i):(v_i,v_j)\in E(G)\}$.
%
%The constrained STG with respect to state \textbf{constraints} $C_\beta^0\subseteq C_\beta$ and
%control constraints $C_u^0(\beta)\subseteq \Delta_M$, $\beta\in C_\beta^0$,
%denoted by $G[C_\beta^0,C_u^0]$, is a
%directed graph, where each vertex represents a state profile in $C_\beta^0$ and each
%directed edge $(\beta_a,\beta_b)$ represents a constrained one-step state
%profile transition from $\beta_a$ to $\beta_b$. In this part, let $\beta_a:=\delta_N^{a}$.
%Then, it holds that
%$V(G[C_\beta^0,C_u^0])=C_\beta^0$ and
%\be\nonumber
%E(G[C_\beta^0,C_u^0])=\{(\beta_a,\beta_b)\in C_\beta^0\times C_\beta^0: \beta_b\in\mathcal{R}_1(\beta_a;C_u^0)\},
%\ee
%where $\mathcal{R}_1(\beta_a;C_u^0)$ denotes the one-step reachable set of state profile
%$\beta_a$ with respect to control constraints $C_u^0(\beta)$ defined by
%\begin{align}\nonumber
%\mathcal{R}_1(\beta_a;C_u^0):=&\{Fu\delta_N^{a}: u\in C_u^0(\delta_N^{a})\}\\\nonumber
%=&\{Col_{a}(Blk_l(F)):\delta_M^l\in C_u^0(\delta_N^{a})\}.
%\end{align}

%Firstly, we study the construction of $I(\Omega(s))$.
Theorem \ref{th01} establishes the equivalence between the
$(Q,\rho)$-asymptotically stability of WCS (\ref{eq01}) and the constrained set stabilization
of FFN (\ref{eq015}). In this part, we extend the approach in \cite{gao2021}
to the constrained set stabilization of FFNs, and the extensions are mainly two-fold.
On one hand, the state constraint and state-dependent control constraint is
considered in this paper. On the other hand, the constrained set stabilization considered in
this paper is originated from the constrained control-state set stabilization, which leads to
the set $I(\Omega(s))$ being an LCCIS, while the key to the set stabilization considered in
\cite{gao2021} is the LCIS contained in a given set.
%the construction of which is different
%from LCCIS (see Remark \ref{rem01}).
%the difference
%between $\tilde{I}(\mathcal{N})$ and $\bar{I}(\mathcal{M})$ lies in that
%$\tilde{I}(\mathcal{N})$ is the largest set of state profiles that can remain in this set
%under a control $u\in\Delta_N$, while $\bar{I}(\mathcal{M})$ is the largest set of state
%profiles $\beta$ that can remain in this set under a control $u\in C_u^{\mathcal{M}}(\beta)$,
%that is,
%$\bar{I}(\mathcal{M})$ is an LCIS contained in $\Phi(\mathcal{M})$ with control constraints
%$C_u^{\mathcal{M}}(\beta)$, $\beta\in\Phi(\mathcal{M})$.
%Thus,
%it is easy to see that $\bar{I}(\mathcal{M})\subseteq\tilde{I}(\Phi(\mathcal{M}))$.}
%Based on the connection between the
%control invariant sets and the strongly connected components of the STG, Gao et al.
%\cite{gao2021} proposed an
%efficient algorithm to obtain the LCIS contained in a given set for Boolean control networks.
By Definition \ref{def03}, one can see that the LCCIS $I(\Omega(s))$ is essentially an LCIS
contained in $\Phi(\Omega(s))$ with control
constraints $C_u^{\Omega(s)}(\beta)$, $\beta\in\Phi(\Omega(s))$. Thus, by virtue of Algorithm 2 in \cite{gao2021}
with replacing $G[\mathcal{M}]$ by $G[\Phi(\Omega(s)),C_u^{\Omega(s)}]$, one can obtain
$I(\Omega(s))$.
%In addition, if $I(\Omega(s))\neq\emptyset$, then for any
%$\delta_N^{a}\in I(\Omega(s))$, the set of control profiles $u\in C_u^{\Omega(s)}(\delta_N^{a})$
%such that $\beta(1;\delta_N^{a},u)\in I(\Omega(s))$ can be given as follows:
%\be\label{eq05}
%U_{a}:=\{\delta_M^l\in C_u^{\Omega(s)}(\delta_N^{a}):\sum_{\delta_N^b\in I(\Omega(s))}Row_b(Col_a(Blk_l(F)))=1\}.
%\ee

Assume that $I(\Omega(s))\neq\emptyset$.
%Since $I(\Omega(s))$ is a CCIS, FFN (\ref{eq015})
%is constrained $I(\Omega(s))$-stabilizable if and only if there exists a state feedback law
%$\pi\in\Pi$ under which all state profiles in $C_\beta\setminus I(\Omega(s))$ can be driven
%to $I(\Omega(s))$. With this in mind,
We construct a STG $G_c$ to check the
constrained $I(\Omega(s))$-stabilization by contracting $I(\Omega(s))$ to a single vertex
$v_{0}$ and
reversing all edges in $G[(C_\beta\setminus I(\Omega(s)))\cup\{v_{0}\},C_u]$.
More specifically, the STG $G_c$ is defined as
\be\label{eq051}
V(G_c)=(C_\beta\setminus I(\Omega(s)))\cup\{v_{0}\}
\ee
and
\begin{align}\nonumber
&E(G_c):=E^\top(G)\cup\{(v_{0},v_{a}): \hbox{there exists}\ \delta_N^{b}\in I(\Omega(s))\\\label{eq052}
&~~~~~~~~~~~\hbox{such that}\ \delta_N^{b}\in\mathcal{R}_1(\delta_N^{a};C_u), v_{a}\in V(G)\},
\end{align}
where $v_{0}$ is a new node representing $I(\Omega(s))$,
$E^\top(G):=\{(v_{b},v_{a}):(v_{a},v_{b})\in E(G)\}$ and
$G:=G[C_\beta\setminus I(\Omega(s)),C_u]$.
%{\definition\label{def010} A STG $G_c$ is called the set stabilization-check STG of
%FFN (\ref{eq015}) with respect to $I(\Omega(s))$,
%if $V(G_c)=(C_\beta\setminus I(\Omega(s))\cup\{\beta_0\}$, and
%\begin{eqnarray}\nonumber
%&&E(G_c):=E^\top(G)\cup\{(\beta_0,\beta_a): \beta_a\in V(G), \hbox{and there}\\\nonumber
%&&~~~~~~~~\hbox{exists}\ \beta\in I(\Omega(s))\ \hbox{such that}\ \beta\in\mathcal{R}_1(\beta_a;C_u)\},
%\end{eqnarray}
%where $\beta_0$ is a new node represents $I(\Omega(s))$ and
%$G:=G[C_\beta\setminus I(\Omega(s)),C_u]$.}
%BFS is a fundamental search algorithm in graph theory. Given a directed graph $G$ and a
%vertex $v_0\in V(G)$, BFS produces a breadth-first spanning tree $T_{v_0}$ rooted
%at $v_0$, where $V(T_{v_0})$ contains all reachable vertices from $v_0$ and
%the path from $v_0$ to any $v\in V(T_{v_0})$ is the shortest one from
%$v_0$ to $v$ in $G$ \cite{cormen2009}.
%In the following, we adopt the breath-first search algorithm \cite{cormen2009} to check
%$(Q,\rho)$-asymptotically stability of WCS (\ref{eq01}).
%By the construction of STG $G_c$, there exists
%a state feedback law
%$\pi\in\Pi$ under which all state profiles in $C_\beta\setminus I(\Omega(s))$ can be driven
%to $I(\Omega(s))$, if and only of $V(T_{\beta_I})=V(G_c)$, where $T_{\beta_I}$ represents
%the breadth-first spanning tree of $G_c$ rooted at $\beta_I$.
Then, we can adopt the breath-first search algorithm \cite{cormen2009} to check the
constrained $I(\Omega(s))$-stabilization of FFN (\ref{eq015}).
%Based on the above analysis, we have the following graph-based criterion for the $(Q,\rho)$-asymptotically stability
%of WCS (\ref{eq01}) taking into account of the WCS and MAS coupling.
%
%{\theorem\label{th011} WCS (\ref{eq01}) is $(Q,\rho)$-asymptotically stable with respect
%to FFN (\ref{eq02}), if and only if $V(T_{\beta_I})=V(G_c)$, where $G_c$ is defined by
%(\ref{eq051}) and (\ref{eq052}),
%and $T_{\beta_I}$ is the
%breadth-first spanning tree of $G_c$ rooted at $\beta_I$.}
%
%\proof\ The result is directly obtained by the equivalence between the
%$(Q,\rho)$-asymptotically stability of
%WCS (\ref{eq01}) and the constrained $I(\Omega(s))$-stabilization of FFN (\ref{eq02}), so
%we omit the proof.
%$\Box$
%Denote the breadth-first spanning tree of $G_c$ rooted at $v_0$ by $T_{v_0}$.
%In $T_{v_{0}}$, denote the set of all paths from
%$v_{0}$ to $v_{a}$ by $P_{v_{0},v_{a}}$, and denote the number of vertices in each path
%$p\in P_{v_{0},v_{a}}$ by $\iota(v_{0},v_{a})$. For any $\delta_N^{a}\in C_\beta\setminus I(\Omega(s))$, define
%\begin{align}\nonumber
%&U_a:=\{\delta_M^l\in C_u(\delta_N^a):Col_a(Blk_l(F))=p[\iota(v_{0},v_a)-1],\\\label{eq07}
%&~~~~~~~~~p\in P(v_{0},v_a)\},
%\end{align}
%where $p[j]$ denotes the $j$-th element of path $p$.

We have the following criterion on
$(Q,\rho)$-asymptotically stability of WCS (\ref{eq01}).
%Then, we have the following result on
%the construction of state feedback laws for FFN (\ref{eq02}) ensuring the
%$(Q,\rho)$-asymptotically stability of WCS (\ref{eq01}).

{\theorem\label{th011} WCS (\ref{eq01}) is $(Q,\rho)$-asymptotically stable with respect
to FFN (\ref{eq015}), if and only if $V(T_{v_0})=V(G_c)$,
where $G_c$ is defined by
(\ref{eq051}) and (\ref{eq052}),
and $T_{v_{0}}$ is the breadth-first spanning tree of $G_c$ rooted at $v_{0}$.
%If WCS (\ref{eq01}) is $(Q,\rho)$-asymptotically stable with respect
%to FFN (\ref{eq02}), then
Moreover, feasible state feedback laws $\pi^\ast$ with state feedback
gain matrices $L^\ast=[l_1^\ast~l_2^\ast~\cdots~l_{N}^\ast]$ can be constructed as follows:
\be\label{eq05}
l_a^\ast\in\left\{
             \begin{array}{ll}
               U_a, & \delta_N^a\in I(\Omega(s)); \\
               \bar{U}_a, & \delta_N^a\in C_\beta\setminus I(\Omega(s)); \\
               \Delta_M, & \delta_N^a\in \Delta_N\setminus C_\beta,
             \end{array}
           \right.
\ee
where
\begin{align}\nonumber
&U_{a}=\{\delta_M^l\in C_u^{\Omega(s)}(\delta_N^{a}):\sum_{\delta_N^b\in I(\Omega(s))}Row_b(Col_a(Blk_l(F)))\\\nonumber
&~~~~~~~~=1\},\\\nonumber
&\bar{U}_a=\{\delta_M^l\in C_u(\delta_N^a):Col_a(Blk_l(F))=p[\iota(v_{0},v_a)-1],\\\nonumber
&~~~~~~~~~p\in P(v_{0},v_a)\},
\end{align}
$P_{v_{0},v_{a}}$ denotes the set of all paths from
$v_{0}$ to $v_{a}$ with the number of vertices
in each path $p\in P_{v_{0},v_{a}}$ being $\iota(v_{0},v_{a})$, and $p[j]$ denotes the $j$-th element of path $p$.}
%In addition, if $I(\Omega(s))\neq\emptyset$, then for any
%$\delta_N^{a}\in I(\Omega(s))$, the set of control profiles $u\in C_u^{\Omega(s)}(\delta_N^{a})$
%such that $\beta(1;\delta_N^{a},u)\in I(\Omega(s))$ can be given as follows:
%\be\label{eq05}
%U_{a}:=\{\delta_M^l\in C_u^{\Omega(s)}(\delta_N^{a}):\sum_{\delta_N^b\in I(\Omega(s))}Row_b(Col_a(Blk_l(F)))=1\}.
%\ee
%\be\nonumber
%l_a^\ast\in\left\{
%             \begin{array}{ll}
%               U_a, & \delta_N^a\in C_\beta; \\
%               \Delta_M, & \delta_N^a\in \Delta_N\setminus C_\beta,
%             \end{array}
%           \right.
%\ee
%where $U_a$ is defined in (\ref{eq05}) and (\ref{eq07}).}

\proof\ Since $I(\Omega(s))$ is a CCIS, FFN (\ref{eq015})
is constrained $I(\Omega(s))$-stabilizable if and only if there exists a state feedback law
$\pi\in\Pi$ under which all state profiles in $C_\beta\setminus I(\Omega(s))$ can be driven
to $I(\Omega(s))$, which is equivalent to $V(T_{v_0})=V(G_c)$ according to the
 construction of STG $G_c$. Then, by Theorem \ref{th01}, the criterion is proved.

On one hand, by the construction of $\bar{U}_a$, under the control of
state feedback laws $\pi^\ast$, the state profile of
FFN (\ref{eq015}) from any initial state profile
$\beta_0\in C_\beta\setminus I(\Omega(s))$ can be driven to $I(\Omega(s))$ along with a corresponding path in $T_{v_0}^\top$,
where $T_{v_0}^\top$ represents a graph obtained by reversing the directions of all
edges in $T_{v_0}$. On the other hand, for any $\delta_N^{a}\in I(\Omega(s))$ and any $u\in U_{a}$, it holds that
$\beta(1;\delta_N^{a},u)\in I(\Omega(s))$. Thus, by Theorem \ref{th01}, under the control of
$\pi^\ast$, WCS (\ref{eq01}) is $(Q,\rho)$-asymptotically stable with respect
to FFN (\ref{eq015}).
%$\pi\in\Pi$ under which all state profiles in $C_\beta\setminus I(\Omega(s))$ can be driven
%to $I(\Omega(s))$, if and only of. In addition, according to (\ref{eq05}) and (\ref{eq07}),
%under $\pi^\ast$ with  $L^\ast$, the state profile of FFN (\ref{eq015}) from any initial state profile $\beta_0$ can be
%driven to and finally remain inside $I(\Omega(s))$.
$\Box$

%{\remark\label{rem02} In the past decade, based on algebraic approaches,
%several results on set stabilization are established for
%finite-value systems including Boolean control networks \cite{guo2015}, multivalued logical
%networks \cite{lif2017}, switched
%Boolean control networks \cite{zhangq2021}. Considering the high
%computational complexity of these results, a graph-based approach is proposed for the set
%stabilization of Boolean control networks in \cite{gao2021}, which is more computationally
%economical than the existing results. In this paper, we extend the approach in \cite{gao2021}
%to the constrained set stabilization of FFNs, and the extensions are mainly two-fold.
%On one hand, the state constraint and state-dependent control constraint is
%considered in this paper. On the other hand, the constrained set stabilization considered in
%this paper is originated from the control-state constrained set stabilization, which lead to
%the set $I(\Omega(s))$ being a LCCIS, while the key to the set stabilization considered in
%\cite{gao2021} is the LCIS contained in a given set, the construction of which is different
%from LCCIS (see Remark \ref{rem01}).}

\section{ILLUSTRATIVE EXAMPLE}
Consider a heterogeneous IIoT system consisting of two autonomous assembly arms and an
automated guided vehicles system (AGVS) of three AGVs. When the raw materials of the
production line are about to run out,
the AGVS enter the workshop of autonomous assembly arms. Then, two autonomous assembly arms
load raw materials into the AGVs by exchanging information between AGVs and remote
controllers via wireless networks. Finally, the AGVS transport raw materials to the
corresponding production line.
%In this process, the physical motion of AGVs may lead to
%shadow fading in the wireless network that the autonomous assembly arms \textbf{are} controlled over.

In the AGVS, assume that only two state agents are responsible for transporting raw
materials, and the control agent does not have the transport ability. Discretize the
workshop of AGVS into three two dimensional regions. Assume that the AGVS has three modes as
\begin{align}\nonumber
&A_{11}=[1~0~0],~A_{12}=[1~0~0],~A_{21}=[1~0~0],\\\nonumber
&A_{22}=[1~0~0];~B_{11}=[1~0~0],~B_{21}=[1~0~0],
\end{align}
where $A_{i,j}$ and $B_{i,l}$ are defined in (\ref{eq012}).
Then, the AGVS is modeled as the following FFN with profile-dependent
switching topology:
\begin{align}\nonumber
&\beta_i(k+1)=a_{i,1}^{\sigma(k)}\times_{3}\beta_1(k)+_{3}a_{i,2}^{\sigma(k)}\times_{3}\beta_2(k)\\\label{eq028}
&~~~~~~~~~~~~~+_{3} b_{i,1}^{\sigma(k)}\times_{3}u_1(k),~i=1,2,
\end{align}
where $u_1(k),\beta_i(k)\in\mathcal{D}_3$, $i=1,2$.

Assume that the two autonomous assembly arms are located in region $0$ and region $2$,
respectively. The dynamics of the WCS of the two autonomous assembly arms is in the form of
(\ref{eq01}) with states $x_1(k)\in\mathbb{R}$ and $x_2(k)=[x_{21}(k)~x_{22}(k)]^\top\in\mathbb{R}^2$
denoting the differences between current and desired states,
%Assume that the dynamics of two autonomous assembly arms are given by
where $A_{c,1}=0.4$, $A_{o,1}=1.1$ and
$$A_{c,2}=\left[
            \begin{array}{cc}
              -0.4 & -0.1 \\
              0.1 & 0.6 \\
            \end{array}
          \right],~
A_{o,2}=\left[
            \begin{array}{cc}
              -1 & -0.4 \\
              -0.5 & 0.3 \\
            \end{array}
          \right].$$
They are perturbed by zero-mean unite-variance Gaussian noises.
Let $\rho_1=0.75$, $\rho_2=0.95$, $Q_1=1$ and suppose that $Q_2$ solves
$A_{c,2}^\top Q_2A_{c,2}=0.7Q_2+I$. Then, by (\ref{eq053}), we can obtain $s_1=0.44$
and $s_2=0.42$.
\begin{figure}[thpb]
      \centering
%      \framebox{\parbox{3in}{We suggest that you use a text box to insert a graphic (which is ideally a 300 dpi TIFF or EPS file, with all fonts embedded) because, in an document, this method is somewhat more stable than directly inserting a picture.
%}}
      \includegraphics[scale=0.45]{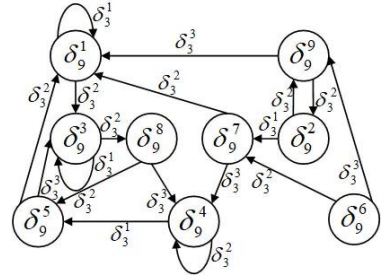}
\vskip-6mm
      \caption{The evolution of the state profiles for AGVS.}
      \label{fig02}
   \end{figure}
The algebraic equivalent form of
profile-dependent switching signal is $\sigma(k)=\Sigma z(k)$,
where $\Sigma=\delta_3[1~1~3~1~1~2~3~2~3~1~1~2~1~1~1~2~1~1~3~2~3~2~1~1~3~1~1]$.
Then, we can obtain the following equivalent algebraic form of FFN (\ref{eq028}):
\be\label{eq090}
\beta(k+1)=Fz(k),
\ee
where $F=\delta_9[1~7~3~5~2~8~2~6~1~3~9~8~4~1~7~1~5~2~6~6~5~1~3~9~4~4$ $1]$.
Let $C_\beta=\Delta_9$ and
$$C_u(\beta)=\left\{
              \begin{array}{ll}
                \{\delta_3^1,\delta_3^2\}, & \beta\in\{\delta_9^i:i=1,\cdots,4\}\\
                \{\delta_3^2,\delta_3^3\}, & \beta\in\{\delta_9^i:i=5,\cdots,9\}.
              \end{array}
            \right.$$
Then, $C_z=\{1,\cdots,4,10,\cdots,18,23,\cdots,27\}$. According to (\ref{eq090}), we can
obtain the evolution of the state profiles for AGVS (Fig. \ref{fig02}).

The WCS and AGVS coupling has equivalent form
$$\mathbb{P}\{\lambda_i(k)=1|z(k)\}=\Lambda_iz(k)$$
with
\begin{align}\nonumber
\Lambda_1=&[0.53~0.21~0.53~0.49~0.21~0.53~0.16~0.00~0.20\\\nonumber
&0.49~0.53~0.21~0.00~0.21~0.53~0.00~0.21~0.53\\\nonumber
&0.00~0.64~0.64~0.00~0.21~0.00~0.00~0.21~0.53],\\\nonumber
\Lambda_2=&[0.67~0.21~0.67~0.00~0.00~0.53~0.00~0.21~0.53\\\nonumber
&0.32~0.67~0.53~0.16~0.00~0.32~0.00~0.32~0.00\\\nonumber
&0.00~0.53~0.16~0.00~0.67~0.53~0.00~0.16~0.00].
\end{align}
%Then, by (\ref{eq050-1}) and (\ref{eq050}), we have
%$\Omega(s_1)=\{\delta_{27}^i:i=1,3,4,10,11,15,18,27\}$,
%$\Omega(s_2)=\{\delta_{27}^i:i=1,3,11,12,23$, $24\}$ and
%$\Omega(s)=\{\delta_{27}^1,\delta_{27}^3,\delta_{27}^{11}\}$.
Then, by (\ref{eq050}), we have
$\Omega(s)=\{\delta_{27}^1,\delta_{27}^3,\delta_{27}^{11}\}$
and $I(\Omega(s))=\{\delta_9^1,\delta_9^3\}$.
Since the two autonomous assembly arms are located in region $\delta_3^1$ and
region $\delta_3^3$, respectively, FFN (\ref{eq028}) accomplishes the transport task
while WCS being $(Q,\rho)$-asymptotically stable with respect to FFN (\ref{eq028}),
if and only if FFN (\ref{eq028}) is constrained
$\bar{I}(\Omega(s))$-stabilizable, where $\bar{I}(\Omega(s)):=\{\delta_9^3\}$.
\begin{figure}[thpb]
      \centering
%      \framebox{\parbox{3in}{We suggest that you use a text box to insert a graphic (which is ideally a 300 dpi TIFF or EPS file, with all fonts embedded) because, in an document, this method is somewhat more stable than directly inserting a picture.
%}}
      \includegraphics[scale=0.42]{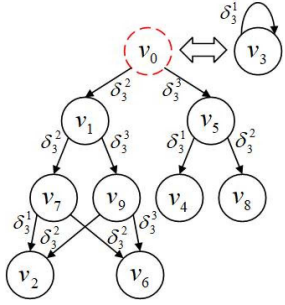}
\vskip-6mm
      \caption{Breadth-first spanning tree $T_{v_0}$ along with $(Q,\rho)$-state
feedback laws.}
      \label{fig03}
   \end{figure}

By constructing the STG $G_c$ defined by (\ref{eq051}) and (\ref{eq052}),
we can draw the conclusion that FFN (\ref{eq028}) is constrained
$\bar{I}(\Omega(s))$-stabilizable. In addition,
according to Theorem \ref{th011}, state feedback gain matrices for feasible state
feedback laws of AGVS can be designed as
\be\label{060}
L^\ast=\delta_3[2~i~1~1~3~j~2~2~3],~i=1,2,~j=2,3.
\ee
The breadth-first spanning tree $T_{v_0}$ along with $(Q,\rho)$-state
feedback laws is shown in Fig. \ref{fig03}.
By recovering the logical forms from algebraic forms (\ref{060}), we finally obtain the
following four state feedback laws for AGVS:
\begin{align}\nonumber
&\pi_1^\ast((0,1))=\pi_2^\ast((0,1))=0,\pi_3^\ast((0,1))=\pi_4^\ast((0,1))=1,\\\nonumber
&\pi_1^\ast((1,2))=\pi_3^\ast((1,2))=1,\pi_2^\ast((1,2))=\pi_4^\ast((1,2))=2,\\\nonumber
&\pi_l^\ast(\beta)=\left\{
               \begin{array}{ll}
                 0, & \beta\in\{(0,2),(1,0)\},\\
                 1, & \beta\in\{(0,0),(2,0),(2,1)\},\\
                 2, & \beta\in\{(1,1),(2,2)\}.
               \end{array}
             \right.~i\in\{1,2,3,4\}.
\end{align}

\begin{figure}[thpb]
      \centering
%      \framebox{\parbox{3in}{We suggest that you use a text box to insert a graphic (which is ideally a 300 dpi TIFF or EPS file, with all fonts embedded) because, in an document, this method is somewhat more stable than directly inserting a picture.
%}}
      \includegraphics[scale=0.5]{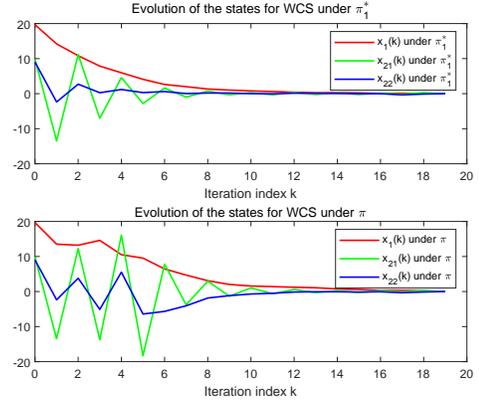}
\vskip-2mm
      \caption{The evolution of the empirical averages of the states
for the WCS under state
feedback law $\pi_1^\ast$ and $\pi$, respectively.}
      \label{fig04}
   \end{figure}

By resorting to the Monte Carlo method and averaging
over $100$ simulation results, we plot in Fig. \ref{fig04} the evolution of the
states
for the WCS with two state agents of the AGVS all
starting from position $2$ under the obtained feasible state
feedback law $\pi_1^\ast$ and a random state
feedback law $\pi$ with $\pi(0,1)=\pi(0,2)=\pi(1,0)=0$, $\pi(0,0)=\pi(1,1)=\pi(1,2)=\pi(2,1)=\pi(2,2)=1$, $\pi(2,0)=2$.
Simulation results show that state feedback laws $\pi_1^\ast$ and $\pi$
all ensure the Lyapunov-like performance of the WCS and the transient period $T$ under
$\pi_1^\ast$ is shorter than that under $\pi$,
%
%satisfied
%from $k_1=1$ and $k_2=$ under $\pi_1^\ast$ while $k_1=1$ and $k_2=$ under $\pi$.
%According to (\ref{eq07}),
and that is because under $\pi_1^\ast$, each state profile of the AGVS enters
$\bar{I}(\Omega(s))$ along with a corresponding path in $T_{v_0}^\top$.
%Since the path from $v_0$ to any vertex $v$
%in $T_{v_0}$ is the shortest one from $v_0$ to $v$ in $G_c$, the state
%feedback law constructed by the method of this paper can achieve faster convergence.
%\textbf{Simulation results show that the expected decay rates $\rho_1$ and $\rho_2$ are
%satisfied
%from $k_1=1$ and $k_2=$ under $\pi_1^\ast$ while $k_1=1$ and $k_2=$ under $\pi$.
%According to (\ref{eq07}), each state profile of the AGVS enters $\bar{I}(\Omega(s))$
%along with a corresponding path in $T_{v_0}$. Since the path from $v_0$ to any vertex $v$
%in $T_{v_0}$ is the shortest one from $v_0$ to $v$ in $G_c$, the state
%feedback law constructed by the method of this paper can achieve faster convergence.}

\section{CONCLUSIONS}
In this paper, we have focused on the controller design for the MAS to ensure the
performance of WCS in the presence of WCS and MAS coupling.
We have adopted the FFN with profile-dependent switching topology to proceed the operational
control for the MAS. Using the ASSR approach, we have obtained an equivalent form
for the WCS and MAS coupling.
Then, we have proposed a criterion in terms of constrained set stabilization
to guarantee the Lyapunov-like performance with expected decay rate.
In addition, based on graph theory and the breath-first searching,
we have designed state feedback controllers for the MAS to ensure the performance
requirements of WCS in the presence of WCS and MAS coupling. Future works will devote to
converting the set stabilization of large-scale MASs into the
set stabilization of subsystems via aggregation method.

%\section*{Acknowledgements}
%The authors would like to thank the Associate Editor and the anonymous reviewers for their constructive comments which improved the quality of this paper.


\begin{thebibliography}{99}     % Otherwise use the
                                 % thebibliography environment.
                                 % Insert the full references here.
                                 % See a recent issue of Automatica
                              % for the style.

\bibitem{fuqaha2015} A. Al-Fuqaha, M. Guizani, M. Mohammadi, M. Aledhari, and
M. Ayyash, Internet of Things: A survey on enabling technologies,
protocols, and applications, {\em IEEE Commun. Surveys Tuts.}, vol. 17, no. 4, pp. 2347-2376,
4th Quart. 2015.

\bibitem{Baumann2021} D. Baumann, F. Mager, U. Wetzker, L. Thiele, M. Zimmerling, S. Trimpe,
Wireless control for smart manufacturing: Recent approaches and open challenges,
{\em Proc. IEEE}, Vol. 109, no. 4, pp. 441-467, Apr. 2021.

\bibitem{fadlullah2011} Z. M. Fadlullah, M. M. Fouda, N. Kato, A. Takeuchi, N. Iwasaki,
and Y. Nozaki, Toward intelligent machine-to-machine communications in smart grid,
{\em IEEE Commun. Mag.}, vol. 49, no. 4, pp. 60-65, Apr. 2011.

\bibitem{vitturi2019} S. Vitturi, C. Zunino, and T. Sauter, Industrial communication systems
and their future challenges: Next-generation Ethernet, IIoT, and 5G, {\em Proc. IEEE}, vol. 107,
no. 6, pp. 944-961, Jun. 2019.



\bibitem{Ahlen2019} A. Ahlen, J. Akerberg, M. Eriksson, A. J. Isaksson, T. Iwaki,
K. H. Johansson, S. Knorn, T. Lindh, and H. Sandberg, Toward wireless control in industrial
process automation: A case study at a paper mill, {\em IEEE Control Syst. Mag.}, vol. 39,
no. 5, pp. 36-57, Oct. 2019.

\bibitem{He2022} W. He, W. Xu, X. Ge, Q. Han, W. Du, and F. Qian, Secure Control of
multiagent systems against malicious attacks: A brief survey, {\em IEEE Trans. Industr.
Inform.}, vol. 18, no. 6, pp. 3595-3608, Jun. 2022.

\bibitem{Pulikottil2021} T. Pulikottil, L. Estrada-Jimenez, H. Rehman, J. Barata,
S. Nikghadam-Hojjati, and L. Zarzycki, Multi-agent based manufacturing: current trends and
challenges, {\em in Proc. 26th IEEE Int. Conf. on Emerging Technologies and Factory
Automation}, Sept. 2021.

\bibitem{liang2019} W. Liang, M. Zheng, J. Zhang, H. Shi, H. Yu, Y. Yang, S. Liu,
W. Yang, and X. Zhao, WIA-FA and its applications to digital factory: A wireless network
solution for factory automation, {\em Proc. IEEE}, vol. 107, no. 6, pp. 1053-1073,
Jun. 2019.

%\bibitem{tanner2003} H. G. Tanner, A. Jadbabaie, and G. J. Pappas, Stable flocking of mobile
%agents, part I: Fixed topology, {\em in Proc. 42nd IEEE Conf. on Decision and Control}, Maui, HI,
%vol. 2, pp. 2010-2015, Dec. 2003.
%
%\bibitem{Choset2005} H. Choset, K. M. Lynch, S. Hutchinson, G. Kantor, W. Burgard,
%and L. E. Kavraki, {\it Principles of Robot Motion: Theory, Algorithms and Implementations.}
%Cambridge, MA, USA: MIT Press, 2005.

\bibitem{Agrawal2014} P. Agrawal, A. Ahl\'{e}n, T. Olofsson, and M. Gidlund, Long term
channel characterization for energy efficient transmission in industrial environments,
{\em IEEE Trans. Commun.}, vol. 62, no. 8, pp. 3004-3014, Aug. 2014.

\bibitem{hu2019} B. Hu, Y. Wang, P. V. Orlik, T. Koike-Akino, and J. Guo, Co-design
of safe and efficient networked control systems in factory automation with state-dependent
wireless fading channels, {\em Automatica}, vol. 105, pp. 334-346, Apr. 2019.

\bibitem{Quevedo2013} D. E. Quevedo, A. Ahlen,  and K. H. Johansson, State estimation
over sensor networks with correlated wireless fading channels, {\em IEEE Trans. Autom.
Control}, vol. 58, no. 3, pp. 581-593, Mar. 2013.

\bibitem{Gatsis2014} K. Gatsis, A. Ribeiro, and G. J. Pappas, Optimal power management
in wireless control systems, {\em IEEE Trans. Autom. Control}, vol. 59, no. 6, pp. 1495-1510,
Jun. 2014.

\bibitem{zhangq1999} Q. Zhang and S. Kassam, Finite-state Markov model for
Rayleigh fading channels, {\em IEEE Trans. Commun.}, vol. 47, no. 11, pp. 1688-1692, Nov. 1999.

\bibitem{gatsis2015} K. Gatsis, M. Pajic, A. Ribeiro, and G. Pappas, Opportunistic
control over shared wireless channels, {\em IEEE Trans. Autom. Control}, vol. 60,
no. 12, pp. 3140-3155, Dec. 2015.

\bibitem{gatsis2018} K. Gatsis, A. Ribeiro, and G. Pappas, Random access design for
wireless control systems, {\em Automatica}, vol. 91, pp. 1-9, Feb. 2018.

\bibitem{Hristu2001} D. Hristu-Varsakelis, Feedback control systems as users of a shared
network: Communication sequences that guarantee stability, {\em in Proc.
40th IEEE Conf. on Decision and Control}, pp. 3631-3636, Dec. 2001.

\bibitem{Molin} A. Molin and S. Hirche, Price-based adaptive scheduling in multi-loop
control systems with resource constraints, {\em IEEE Trans. Autom. Control},
vol. 59, no. 12, pp. 3282-3295, Dec. 2014.

\bibitem{zhang2006} L. Zhang and D. Hristu-Varsakelis, Communication and control co-design for networked control
systems, {\em Automatica}, vol. 42, no. 6, pp. 953-958, Jun. 2006.

\bibitem{le2011} J. Le Ny, E. Feron, and G. J. Pappas, Resource constrained LQR control
under fast sampling, {\em in Proc. 14th Int. Conf. Hybrid Syst.: Comput.
Control}, pp. 271-280, Apr. 2011.

\bibitem{Park2018} P. Park, S. C. Ergen, C. Fischione, C. Lu, and K. H. Johansson, Wireless network
design for control systems: A survey, {\em IEEE Commun. Surv. Tutor.}, vol. 20,
no. 2, pp. 978-1013, Second Quart. 2018.

\bibitem{hu2020} B. Hu, Event-based adaptive power control in vehicular networked systems
with state-dependent bursty fading channels, {\em IEEE Transactions on Circuits and Systems
II: Express Briefs}, Vol. 67, no. 3, pp. 506-510, Mar. 2020.



\bibitem{guo2015} Y. Guo, P. Wang, W. Gui, and C. Yang, Set stability and set stabilization of Boolean control networks based
on invariant subsets, {\em Automatica}, vol. 61, pp. 106-112, Nov. 2015.

\bibitem{lif2017} F. Li, H. Li, L. Xie, and Q. Zhou, On stabilization and set stabilization
of multivalued logical systems, {\em Automatica}, vol. 80, pp. 41-47, Jun. 2017.

\bibitem{zhangq2021} Q. Zhang, J. Feng, Y. Zhao, and J. Zhao, Stabilization and set
stabilization of switched Boolean control networks via flipping mechanism, {\em Nonlinear
Anal-Hybri.}, vol. 41, pp. 101055, May 2021.

\bibitem{cheng2011} D. Cheng and H. Qi, {\it Analysis and Control of Boolean Networks:
A Semi-tensor Product Approach}. London, U.K.: Springer-Verlag, 2011.

\bibitem{Pasqualetti2014} F. Pasqualetti, D. Borra, and F. Bullo, Consensus networks over
finite fields, {\em Automatica}, vol. 50, pp. 349-358, Feb. 2014.

\bibitem{Lidl1996} R. Lidl and H. Niederreiter, {\it Finite Fields}.
New York, NY, USA: Cambridge University Press, 1996.

\bibitem{li2018} H. Li, G. Zhao, P. Guo, and Z. Liu, {\it Analysis and Control of
Finite-Value Systems}. Boca Raton, FL, USA: CRC Press, 2018.

\bibitem{Liy2019} Y. Li, H. Li, X. Ding, and G. Zhao, Leader-follower consensus of
multiagent systems with time delays over finite fields, {\em IEEE Trans. Cybern.},
vol. 49, no. 8, pp. 3203-3208, Aug. 2019.

\bibitem{gao2021} S. Gao, C. Xiang, and T. Lee, Set invariance and optimal
set stabilization of Boolean control networks: A graphical approach, {\em IEEE Trans. Control.
Netw. Syst.}, vol. 8, no. 1, pp. 400-412, Mar. 2021.

\bibitem{cormen2009} T. H. Cormen, C. E. Leiserson, R. L. Rivest, and C. Stein,
{\it Introduction to Algorithms}, 3rd ed. Cambridge, MA, USA: MIT Press, 2009.

%\bibitem{sisinni2018} E. Sisinni, A. Saifullah, S. Han, U. Jennehag, and M. Gidlund,
%Industrial Internet of Things: Challenges, opportunities, and directions, {\em IEEE Trans.
%Industr. Inform.}, vol. 14, no. 11, pp. 4724-4734, Nov. 2018.
%\bibitem{wangx2021} X. Wang, C. Chen, J. He, S. Zhu, and X. Guan, AoI-Aware control and
%communication co-design for industrial IoT systems, {\em IEEE Internet Things J.}, vol. 8, no. 10,
%pp. 8464-8473, May 2021.
%\bibitem{liy2018} Y. Li, B. Li, Y. Liu, J. Lu, Z. Wang, and F. E. Alsaadi, Set stability and
%stabilization of switched Boolean networks with state-based switching,
%IEEE Access, vol. 6, pp. 35624-35630, Jun. 2018.
%\bibitem{Bao2011} L. Bao, M. Skoglund, and K. H. Johansson, Iterative
%encoder-controller design for feedback control over noisy channels, {\em IEEE Trans. Autom.
%Control}, vol. 56, no. 2, pp. 265-278, Feb. 2011.
%\bibitem{Molin2009} A. Molin and S. Hirche, On LQG joint optimal scheduling and
%control under communication constraints, {\em in Proc. 48th IEEE Conf. on
%Decision and Control}, pp. 5832-5838, Dec. 2009.
%\bibitem{goldsmith2005} A. Goldsmith, {\it Wireless Communications}. Cambridge, U.K.:
%Cambridge University Press, 2005.
%\bibitem{tse2005} D. Tse and P. Viswanath, {\it Fundamentals of Wireless Communication}.
%Cambridge, U.K.: Cambridge University Press, 2005.
%\bibitem{vitturi2013} S. Vitturi, F. Tramarin, and L. Seno, Industrial wireless networks:
%The significance of timeliness in communication systems, {\em IEEE Ind. Electron. Mag.},
%vol. 7, no. 2, pp. 40-51, Jun. 2013.
%\bibitem{lu2020} P. Lu, W. Yu, G. Chen, and X. Yu, Leaderless consensus of ring-networked
%mobile robots via distributed saturated control, {\em IEEE Trans. Ind. Electron.},
%vol. 67, no. 12, pp. 10723-10731, Dec. 2020.
%\bibitem{Zavlanos2008} M. M. Zavlanos and G. J. Pappas, Distributed connectivity control of
%mobile networks, {\em IEEE Trans. Robot.}, vol. 24, no. 6, pp. 1416-1428,
%Dec. 2008.
%\bibitem{Mastellone2008} S. Mastellone, D. M. Stipanovic, C. R. Graunke, K. A. Intlekofer,
%and M. W. Spong, Formation control and collision avoidance for
%multi-agent non-holonomic systems: Theory and experiments, {\em The Int.
%Journal Robot. Res.}, vol. 27, no. 1, pp. 107-126, Jan. 2008.
%\bibitem{Fainekos2009} G. E. Fainekos, A. Girard, H. Kress-Gazit, and G. J. Pappas, Temporal
%logic motion planning for dynamic robots, Automatica, vol. 45, no. 2,
%pp. 343-352, 2009.
%\bibitem{liy} Y. Li, H. Li, and X. Ding, Set stability of switched delayed logical networks
%with application to finite-field consensus, Automatica, vol. 113, pp. 108768, Dec. 2020.
%\bibitem{zhangx2020} X. Zhang, Y. Wang, and D. Cheng, Output tracking of Boolean control
%networks, {\em IEEE Trans. Autom. Control}, vol. 65, no. 6, pp. 2730-2735, Jun. 2020.
%\bibitem{liu2020} R. Liu, J. Lu, W. X. Zheng, and J. Kurths, Output feedback control for
%set stabilization of Boolean control networks, {\em IEEE Trans. Neural Netw.
%Learn. Syst.}, vol. 31, no. 6, pp. 2129¨C2139, Jun. 2020.
%\bibitem{Dobslaw2016} F. Dobslaw, T. Zhang, and M. Gidlund, End-to-end reliability-aware
%scheduling for wireless sensor networks,¡± IEEE Trans. Ind. Informat.,
%vol. 12, no. 2, pp. 758-767, Apr. 2016.
\end{thebibliography}
\end{document}